\documentclass[12pt]{article}
\usepackage[T1]{fontenc}
\usepackage{epsfig}
\begin{document}
\newcommand{\ksi}{\vec{\xi}}
\newcommand{\be}{\begin{equation}}
\newcommand{\ee}{\end{equation}}
\newcommand{\beq}{\begin{eqnarray}}
\newcommand{\eeq}{\end{eqnarray}}
\newcommand{\sma}{Sm~A}
\newcommand{\alfa}{Sm~C$^*_\alpha$~}
\newcommand{\fe}{Sm~C$^*$~}
\newcommand{\afe}{Sm~C$^*_A$~}
\newcommand{\heli}{Sm~C$^*_H$~}
\newcommand{\fer}{Sm~C$^*_{FI1}$~}
\newcommand{\ferr}{Sm~C$^*_{FI2}$~}
\newcommand{\rr}{ \stackrel{I}{\longleftrightarrow}}
\newcommand{\rrr}{\stackrel{II}{\longleftrightarrow}}
\newcommand{\rlh}{\longleftrightarrow}
\newcommand{\rla}{$\leftrightarrow$}
~\\\\\\ {\bf Short Pitch Structures in Smectics with
Interactions over more than Two Layers}
\\\\\\
MOJCA \v CEPI\v C$^{a,b}$, BARBARA ROV\v SEK$^{a,c}$,
BO\v STJAN \v ZEK\v S$^{a,d}$ \\
$^{a}$ J. Stefan Institute, Jamova 39;
$^{b}$ Faculty of Education, Kardeljeva pl. 16;
$^c$ Faculty of Mathematics and Physics, Jadranska 19;
$^{d}$ Institute of Biophysics, Medical Faculty, Lipi\v ceva 2,
all 1000 Ljubljana, Slovenia;
\\\\\\
\small
Phenomenological model of chiral polar smectics is introduced
with interactions up
to the fourth neighboring layers. The minimization of the free
energy gives three stable structures: the ferroelectric Sm C$^*$
phase, the antiferroelectric  Sm C$^*_A$ phase and the hellicoidally
modulated structure of the Sm C$^*_H$ phase. The \heli phase
can be recognized
as the Sm C$^*_\alpha$ phase below \sma~ phase and
as the \ferr phase or as the \fer phase  when appearing between
 the \fe phase and the \afe phase.
Stability of these phases is analyzed and the phase
diagram in the space of model parameters is presented.
\\\\
\underline{Keywords:} antiferroelectric liquid crystals;
interlayer interactions; hellicoidal
modulations.\\\\\\
\normalsize
\noindent
{\underline {INTRODUCTION}
\\\\
Recent observation of various subphases in
the thiobenzoate antiferroelectric liquid crystal 10OTBBB1M7
performed by polarized x-ray
scattering [1] has revealed the structure of the two phases which are stable
in the temperature region between the ferroelectric \fe phase and the
antiferroelectric \afe phase. The studied material has the following
phase sequence with decreasing temperature
\sma~ \rla~ \alfa~ \rla~ \fe~ \rla~ \ferr~
\rla~ \fer~ \\
\rla~ \afe~
with first
order transitions between the phases. The phase sequence in this material
is general for antiferroelectric liquid
crystals. In some materials some of the phases are missing [2].

The \alfa phase which appears directly below the \sma~ phase
is hellicoidally modulated over few layers and its period is in general not
commensurate with the layer thickness [3].
The polarized x-ray scattering [1] has
confirmed its structure, which is also consistent with
indirect ellipsometric observations [4,5].

The \ferr phase appears directly below the
\fe phase. It is modulated over
approximately four smectic layers. Only the structures where the molecular
tilts are not in a single plane are consistent with the
experimental data. The simplest structure consistent with experiments
is presented within the clock
model where tilts of the molecules are constant in the magnitude but
differ in direction from the layer to the layer for an approximate
angle $\pm 90^\circ$.
The phase difference has only one sign, and has opposite signs in the
two enantiomers of the same materials. The hellicoidal modulation
in these systems has a very short pitch of approximately four  layers.

The \fer phase which appears below the \ferr phase and above the \afe
phase is hellicoidaly modulated
over approximately three layers. Again, in the most simple structure
the difference of the tilt direction is approximately 120$^\circ$.
 For this phase the typical textures varies with time also in a
temperature stabilized sample.  Also
peaks of the x-ray measurements are much wider than in other phases,
which could be the consequence of various defects or fluctuations.

The structures with short helices were theoretically predicted in systems
with competing interactions between nearest and next nearest
neighboring layers [3]. In the extended model of
antiferroelectric liquid
crystals [6], the \alfa phase was predicted as the
hellicoidally modulated
over few layers only, while the structure between the \fe and the \afe phase
consisted of two helices geared into each other for a general angle.
The last structure although consistent with various observations
[2], has
been ruled out by the first direct structural observation [1].

In this paper we analyze the consequences of the interactions which
extend over more than two neighboring layers. In order to stabilize
structures with three and four layers periodicities, coupling terms
of the same range are needed [7]. Therefore we introduce such
terms in the
expression for the free energy and analyze their influence on the various
structures in these systems. We find three different stable structures:
the structure with the synclinic tilt or the \fe phase, the structure with
the anticlinic tilt or the \afe phase and the hellicoidally modulated
structure
we called the Sm C$^*_H$ phase. The \heli phase with the phase
difference $\alpha$ approximatelly 90$^\circ$ is stable when
interactions between nearest neighboring layers are weak and interactions
with next nearest neighboring
layers favor anticlinic ordering [3]. The structure corresponds to the
\ferr phase. The \heli phase with the modulation of approximately
three layers corresponds to the \fer phase and is stabilized by the
interactions with third neighboring layers which favor synclinic tilts.
In addition we find that interactions with third
and fourth neighboring layers cause first order transition between the
\fe and the \heli phase and between the \afe and the \heli phase,
and can account for the first order transitions between the \fer
phase and the \afe phase or between the \fer phase and the \fe
phase. For some sets of model parameters
two \heli phases with two different periods are
stable and can account for the first order transition between the \fer and
the \ferr phase. The results are given in phase diagrams in the
model parameter space. Finally, we conclude and discuss some open
questions.
\\
\\
\noindent
{\underline {FREE ENERGY }
\\\\
In the system where periodical structures with periods over three or
four smectic layers exist, we expect some interactions
over the same number of layers. Although it is not reasonable to
expect that direct interactions extend over more than two layers
layers [8], indirect interactions of longer range are possible
due to the flexoelectric interactions [9]. To describe effective
direct and
indirect interactions we write tilt order parameter in the $j$-th layer
as $ \vec{\xi}_j = <\{ n_{j,x} n_{j,z}, n_{j,y} n_{j,z}\}> =
\{\xi_{j,x}, \xi_{j,y} \}$.
Achiral interlayer interactions in their simplest form are given as
scalar products between tilt vectors in neighboring layers weighted by
the model parameters which give the strength of the interlayer coupling.
In this analysis chiral interactions will be  neglected. The free energy
of the system with interactions up to fourth neighboring layers
\begin{equation}
G = G_0 + \sum_j
a_1 \left ( \vec{\xi}_j \cdot \vec{\xi}_{j+1} \right ) +
a_2 \left ( \vec{\xi}_j \cdot \vec{\xi}_{j+2} \right ) +
a_3 \left ( \vec{\xi}_j \cdot \vec{\xi}_{j+3} \right ) +
a_4 \left ( \vec{\xi}_j \cdot \vec{\xi}_{j+4} \right ).
\label{g0}
\end{equation}
In the expression above $G_0$ is the part of the free energy which
does not depend of the interlayer interactions. Model parameters $a_1$ to
$a_4$ give the strength of interlayer interactions and favor synclinic
tilts in interacting layers for their negative signs and anticlinic
tilts for their positive signs [3]. Interactions are in general
competing.
The sign of $a_1$ can be either positive or negative  [8,10].
In this analysis the sign of the $a_2$
parameter will be
taken as positive since this is expected for the systems with hellicoidal
modulations over few layers. To stabilize the structure with the three
layer periodicity the sign of the $a_3$ has to be negative, the fact
which is also in agreement with the physical origin of this interaction
[11]. The four layer structures are stabilized already for the
significant $a_2$ model parameter, therefore we consider both signs of
the model parameter $a_4$.

\begin{figure}[th]
\begin{center}\epsfig{file=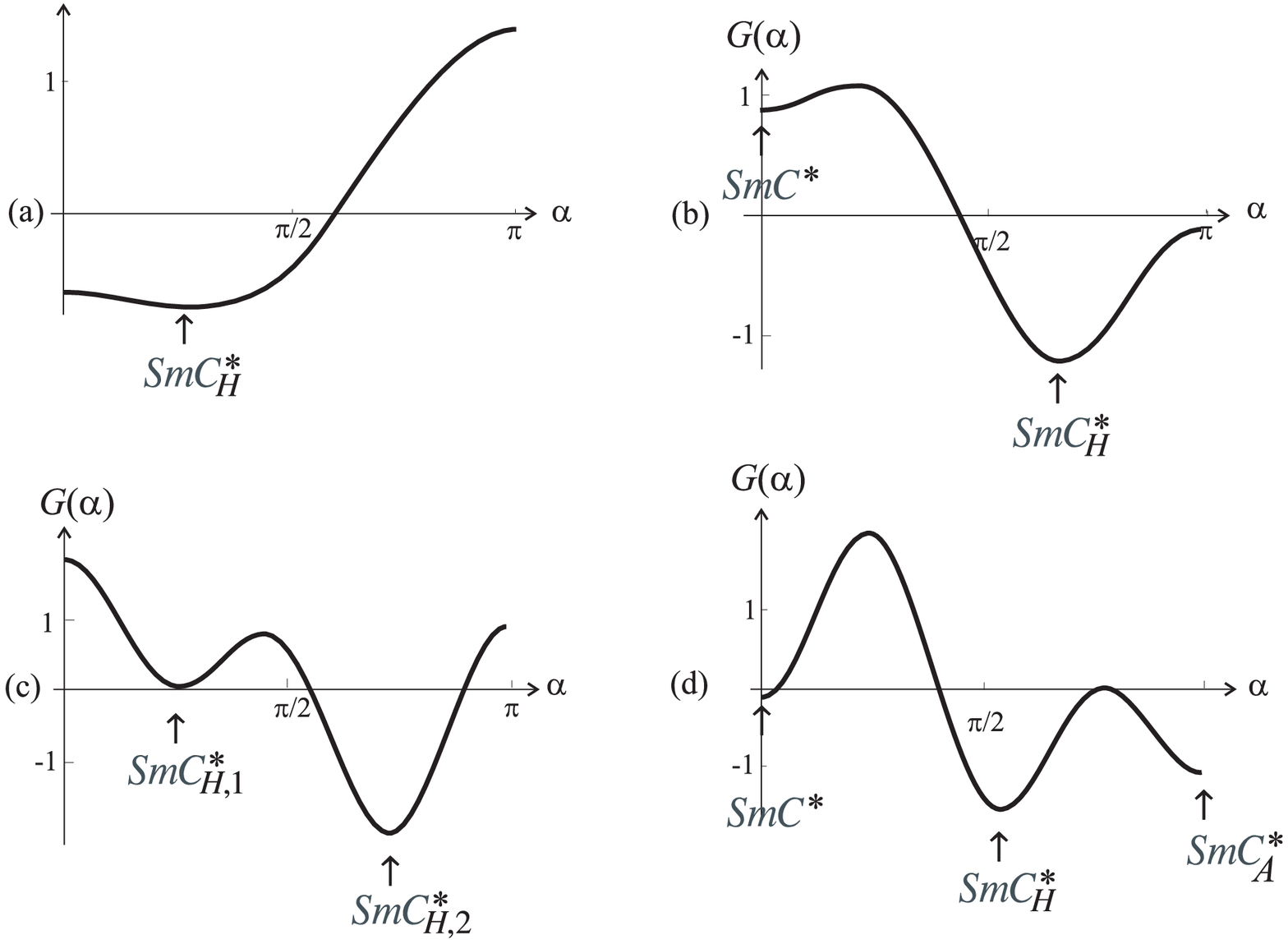,width=120mm}\end{center}
\caption{\small Free energy $G(\alpha)$
for the model parameters:
a) $a_1 = -1.0, a_2 = 0.4, a_3 = 0.0, a_4 = 0.0$;
one possible stable phase,
b) $a_1 = 1.0, a_2 =0.4 , a_3 = -0.5, a_4 = 0.0 $;
two possible stable phases,
c) $a_1 = 1.0, a_2 =0.4 , a_3 =-0.5, a_4 = 1 $;
two possible stable \heli phases,
d) $a_1 = 1.0, a_2 =0.4 , a_3 =-0.5, a_4 = -1 $;
three possible stable phases.}
\end{figure}
To minimize the free energy (Eq.~\ref{g0}) we assume that the tilt varies
only in its direction and not in its magnitude and is given by $\theta$.
Further, here we analyze
only structures of the clock model type, which means that the difference
of the tilt directions in neighboring layers $\alpha$ is constant in the
magnitude and the sign. Such structures we called the \heli
phase for a general
value of $\alpha$ different from 0 or $\pi$.
We therefore look for the solution of the type
\begin{equation}
\vec{\xi}_j = \theta \{ \cos j \alpha, \sin j \alpha \}
\end{equation}
where the initial phase angle in the first layer is set to zero which is
allowed due to the
rotational symmetry. The free energy has now the form
\begin{equation}
G = G_0 + \theta^2\: \left (
a_1 \;\cos\alpha +
a_2 \;\cos2\alpha +
a_3 \;\cos3\alpha +
a_4 \;\cos4\alpha
\right ).
\label{alfa}
\end{equation}
Since we are interested only in the periods of structures and not in the
tilt magnitude, we have to minimize the free energy only with respect to
the phase difference $\alpha$. Derivation with respect to $\alpha$ gives
\begin{equation}
\left(a_1 - 3 a_3 + 4(a_2 - 4 a_4) \cos \alpha + 12 a_3 \cos^2 \alpha +
32 a_4 \cos^3 \alpha \right ) \sin \alpha =0.
\label{cos}
\end{equation}
The solutions of the (Eq.~\ref{cos}) present locally stable minima providing
second derivative with respect to $\alpha$ is positive. For
different sets of model parameters we find one, two or three
locally stable structures.

 In Fig. 1 we present free energy $G(\alpha)$ for a few sets of model
parameters. In systems with only nearest and next nearest neighbors
interactions, only one stable solution for the phase difference
$\alpha$ is possible (Fig. 1a) and corresponds to the \fe phase, the
\afe phase or the \heli phase. In systems where additional third
nearest neighbors interactions are not negligible, the \fe phase and the
\heli phase can be stable for the same set of
parameters, which gives rise to the first order transition between
the two of them (Fig. 1b). If also fourth nearest neighbors
interactions are present, they can stabilize either two \heli
 modulated phases (Fig. 1c) or even three phases - the
\fe phase, the \afe phase and the \heli phase for the same set of model
parameters (Fig. 1d).
\newpage
\noindent
{\underline {PHASE DIAGRAM AND THE STABILITY OF THE PHASES}
\\[6pt]
From the symmetry point of view, all structures with various values of
the phase difference $\alpha$ are the same and can be continuously
transformed. There always exists a
symmetry operation which consists of the translation for the layer thickness
along the
layer normal combined with the rotation for a general angle $\alpha$
around the layer normal. The structural parameter phase angle
$\alpha$ can  be used as a typical parameter which describes the
structure. Its value
also strongly affects the macroscopic properties of the samples.
Since in less recent experiments [2] only macroscopic properties of the
materials have been measured, that observations lead to the names of the
phases according to their macroscopic properties. Recent
experiments [1]
 have revealed that some phase angles $\alpha$ are typical
 for these phases, see Table 1.
\begin{table}
\begin{tabular}{|c|c|c|}
\hline
{\bf the phase} & {\bf phase angle} & {\bf properties}\\
\hline
SmA & not defined & paraelectric, no ORP\\
\hline
SmC$^*$& $\approx 0$ & ferroelectric, ORP\\
\hline
SmC$^*_A$& $\approx \pi$ & antiferroelectric, ORP\\
\hline
SmC$^*_\alpha$ & general value & mainly paraelectric, no ORP\\
\hline
\ferr & $\approx \pi/2$ & antiferroelectric, ORP\\
\hline
\fer & $\approx 2\pi/3$ & ferrielectric, ORP\\
\hline
\end{tabular}
\caption{\small Phases and corresponding phase angles $\alpha$. The value of
the phase angle is not exactly 0 or $\pi$ in chiral samples. Some
macroscopic properties are also given.}
\end{table}
Temperatures of the phase transitions were measured by
differential scanning calorimetry which can detect
only first order
transitions. Sometimes the temperatures are defined through changes of
the texture observations and/or drastical changes of dielectric
properties, optical rotatory power (ORP) and others. The transition
temperatures are in all
mentioned cases more or less unprecise due to the hysteresis or due to
the arbitrary decision which order of the certain property is strong
enough to define the phase. We
have to mention again that all transitions are isostructural and therefore
the transition temperatures can be observed only for the first order
transitions.

\begin{figure}[ht]
\begin{center}\epsfig{file=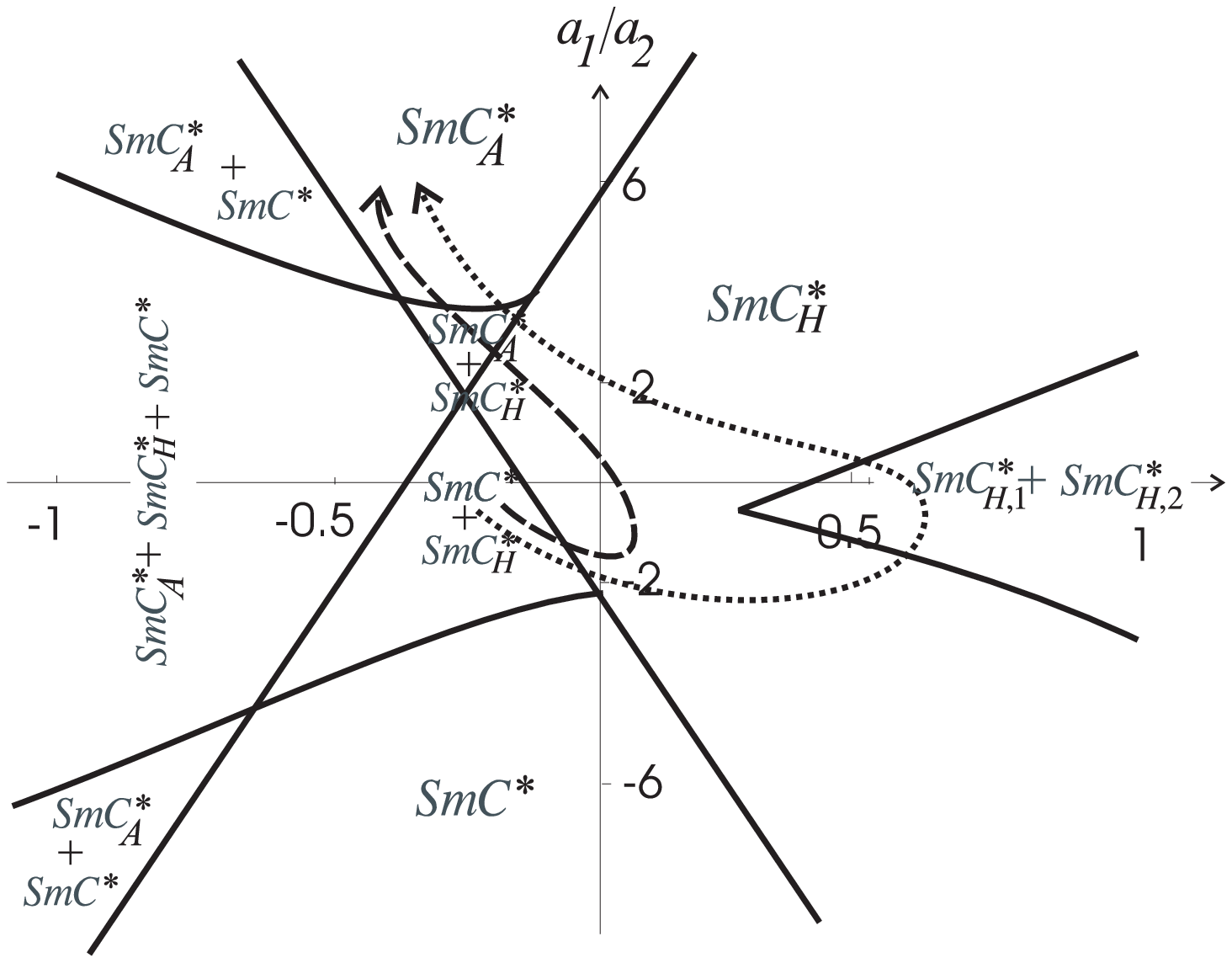,width=120mm}\end{center}
\caption{\small Phase diagram for the ratio $|a_3|/a_2 = 0.2$.
A tentative path in the space of model parameters for MHPOBC (dashed line)
and the 10OTBBB1M7 (doted line)is given. The arrow marks
decreasing temperature.}
\end{figure}
\begin{figure}[ht]
\begin{center}\epsfig{file=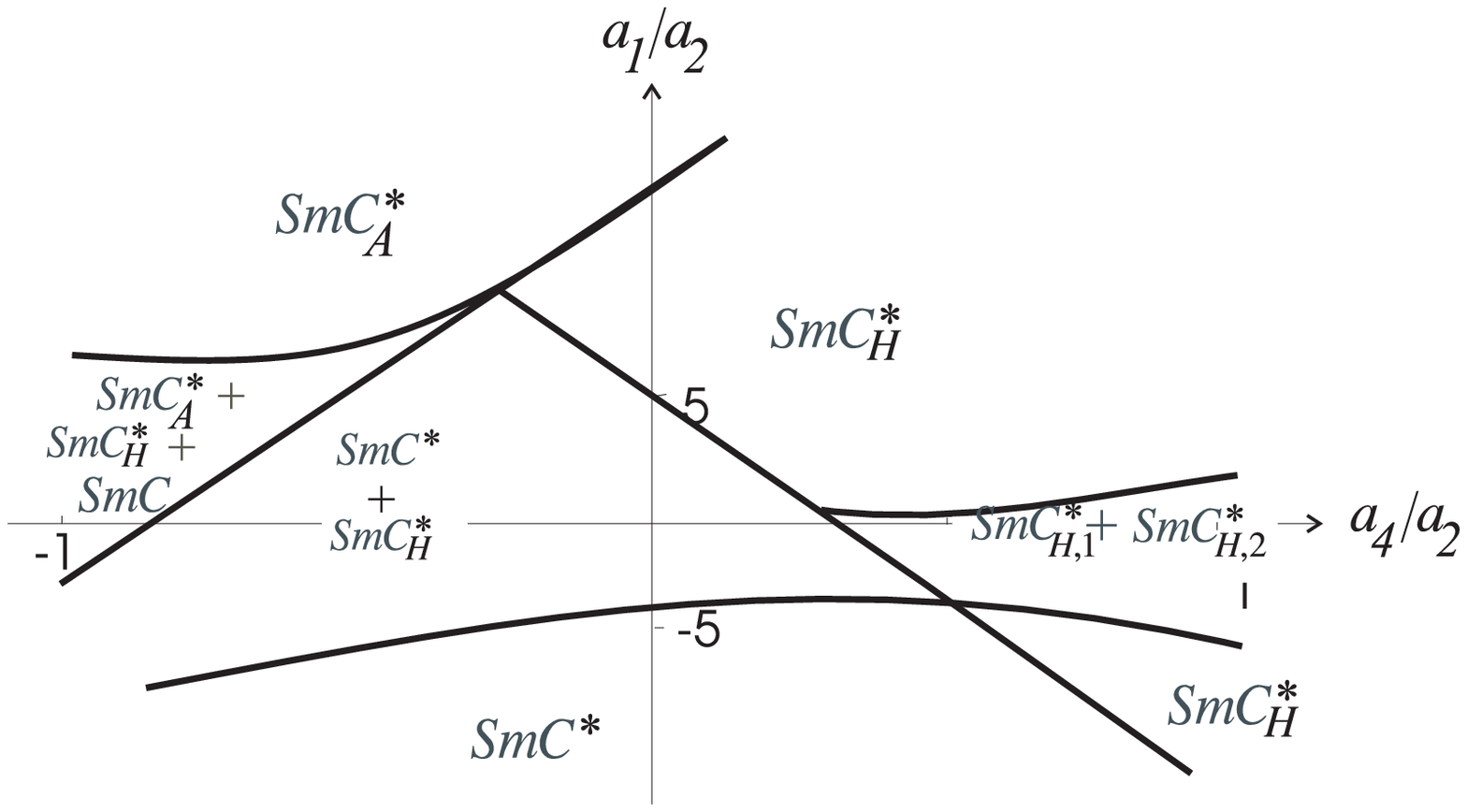,width=120mm}\end{center}
\caption{\small Phase diagram for the ratio $|a_3|/a_2 = 1$. }
\end{figure}

In the phase diagram (Fig. 2 and 3) we therefore did not distinguish
between phases with general angle $\alpha$ or the \heli phase and
phases,
where the phase angle $\alpha$ is approximately $\pi/2$ or $2\pi/3$,
the structures which presumably correspond to the \ferr and the \fer phase,
respectively. We
used for the ferroelectric \fe phase and the antiferroelectric \afe phase
their historical name.

All model parameters change with temperature, since they are the
consequence of microscopic interactions which can also change with the
temperature because of the changing nematic, smectic and tilt order of
the sample as well as the entropical disorder.

The phase diagram of various structures is given for two different
ratios of $a_3/a_2$. The analysis is made only for the competing next
nearest neighbors interactions or $a_2>0$ and the values of model
parameters are expressed in units of $a_2$. The value of parameter $a_3$
is chosen as a parameter of the phase diagram. The regions where one or
more phases are stable, are separated by lines. The stable phases are
marked in the region according to the previous discussion. In the
region, where the solutions for the two different phase angles $\alpha$
are stable at the same temperatures are marked as the
Sm~C$^*_{H,1}$ and Sm~C$^*_{H,2}$.

The third nearest neighbors coupling is relatively weak in Fig. 2 and
much stronger in Fig. 3. As known also from other systems, the phase
diagram for stronger coupling (Fig. 3) is more simple. In general, for
strong nearest layers couplings $|a_1/a_2|$, larger than other couplings,
system always favor either ferroelectric or antiferroelectric ordering
regions of the single stable phase (\fe or \afe). Third nearest
neighbors coupling always favors synclinic third nearest neighbors and
therefore encourage either simple synclinic ordering or the three layer
structures. It therefore introduces first order transition from the
ferroelectric \fe phase to the \heli phase as well as broaden the
region of the stable \heli phases with pitches close to three layers.
The result is seen as the expansion of the two phase region (\fe and
\heli) and the shrinking of the other two phase region (\afe and \heli).
The fourth nearest neighbors interactions favor synclinic ordering for
its negative sign and therefore favor in the same way \fe, \afe as well
as the four layers \heli. For its large negative values all
three phases can
be stable for the same temperature.  For its positive value
it encourages the competition between various interlayer interactions and
gives rise to two stable \heli phases at the same temperature.

To simulate the behaviour of the most studied antiferroelectric liquid
crystal MHPOBC, the parameter changes can follow the path in the
parameter space as marked by the dashed line on Fig. 2. The path direction
marked by an arrow follows the decreasing of the temperature.
When the system
becomes tilted, the \heli phase is stable. It has a general value od the
phase difference $\alpha$ and as such is recognized as the \alfa phase.
 With lowering temperature,
parameters changes their values and the \fe phase becomes stable by the first
order transition. By cooling again the \heli phase with approximately
three layers stabilizes, which corresponds to the \fer phase and
finally the \afe phase becomes stable
by the first order transition. The parameter path of the material
10OTBBB1M7
studied by x-ray measurements has to go through region of the two stable
\heli phases (dotted line), since two hellicoidally modulated phases
separated by the first order transition exist. We have to add that such
a parameter path is
only tentative, since in real system we have to expect, that it goes
through different planes of the 3D space of model parameters ($a_1/a_2,
a_3/a_2, a_4/a_2$). To find the real path, also tilt dependence as well
as precise phase angle dependence versus temperature, has to be known.
~\\
\\
\noindent
{\underline {CONCLUSIONS}}\\[6pt]
In this paper we present the phenomenological model with interactions up
to the fourth nearest neighbors. We analyze the local stabilities of the
synclinic or the ferroelectric \fe phase, the anticlinic or the \afe
phase and the hellicoidal structure with short and in general
incommensurate pitch or the \heli phase. All structures are in
principle of the same symmetry which is defined by symmetry operation
consisting from the translation along the layer normal for a layer thickness
and rotation for an angle $\alpha$ which can have any value between
$-\pi$ and $\pi$. For some special values of the phase
difference $\alpha$, the structures has already been known by decades.
The phase difference has $\alpha =0$ for the ferroelectric \fe phase,
$\alpha =\pi$ for the antiferroelectric phase and has general values in
the \alfa phase. In the \ferr and the \fer phases, the angles are
approximately $\pi/2$ and $2\pi/3$, respectively. The last two
structures can therefore be recognized as special cases of the general
\heli structure.

Interactions of longer range induce in the system first order
transitions between phases. By the first order transitions, transition
temperatures between different phases become observable, since all the
transitions are of the isostructural type. We also discuss the
arbitrariness of the transition temperatures defined by observations of
macroscopic properties, when transitions are continuous.

In the phase diagram the stability regions of different phases in the
model parameters space are presented. Tentative temperature dependence
of model parameters is also shown for the two mostly studied
materials, MHPOBC and 10OTBBB1M7.

However, in this article only the consequences of interactions of longer
range are discussed. Their physical origin is presented in the separate
paper [11], where also the physically reasonable magnitudes of the
parameters
are considered. Within the stability analysis only clock model has been
taken into account and possible variations of the phase differences has
not been considered and remain as a future problem.
\\
\noindent
{\underline {ACKNOWLEDGEMENT}}\\
The financial support of
the Ministry of Science of  Republic Slovenia is
acknowledged.
\\
\\
{\bf References}}\\
~[1] P. Mach, R. Pindak, A.M. Levelut, P. Barois,H.T. Nguyen, C.C. Huang,
L. Furenlid, \underline{ PRL 81}, 1105, (1998).\\
~[2] A. Fukuda, Y. Takanishi, T. Isozaki, K. Ishikawa, H. Takezoe,
\underline{J. Mater. Chem. 4}, 997, (1994) and references therein.\\
~[3] M. \v Cepi\v c, B. \v Zek\v s, \underline{Mol. Cryst. Liq.
Cryst.},\underline{263}, 61 (1995).\\
~[4] Ch. Bahr, D. Fliegner, C.J. Booth, J.W. Goodby, \underline{PRE 51},
R3823 (1995).\\
~[5] B. Rov\v sek, M. \v Cepi\v c, B. \v Zek\v s, \underline{PRE 54},
R3113 (1996).\\
~[6] M. \v Skarabot, M. \v Cepi\v c, B. \v Zek\v s, I. Mu\v sevi\v c, G.
Heppke, A.V.Kityk, I. Mu\v sevi\v c,
\underline{Phys.Rev.E}., \underline{58}, 575, (1998).\\
~[7] A. Roy, N.V. Madhusudana, \underline{Eur. Phys. J. E, 1}, 329 (2000).\\
~[8] M. \v Cepi\v c, B. \v Zek\v s, \underline{Mol. Cryst. Liq.
Cryst.},\underline{301}, 221, (1997).\\
~[9] M. \v Cepi\v c, B. Rov\v sek, B. \v Zek\v s, to appear in
Ferroelectrics.\\
~[10] B. \v Zek\v s, M. \v Cepi\v c, \underline{Proc. of SPIE 3318},
68 (1998).\\
~[11] M. \v Cepi\v c, B. Rov\v sek, B. \v Zek\v s, submitted for
publication.
\end{document}